\documentclass[3p,times,twocolumn]{elsarticle}

\usepackage{natbib}

\usepackage{amssymb}

\usepackage[figuresright]{rotating}
\usepackage{epsfig}
\usepackage{graphicx,setspace}
\usepackage{color}
\usepackage{epstopdf}
\usepackage{natbib}

\begin{document}

\begin{frontmatter}




\title{Collectivity in Small QCD Systems}


\author{Theodore Koblesky on behalf of the PHENIX Collaboration}

\address{Department of Physics, University of Colorado, Boulder, CO 80309, USA}

\begin{abstract}
Features of collectivity observed in high energy A+A collision systems at the Relativistic Heavy Ion Collider (RHIC) and the Large Hadron Collider (LHC) imply a strongly coupled quark gluon plasma (QGP) that flows. One defining feature of collectivity is long-range angular correlations which are characterized experimentally by measuring the azimuthal anisotropy of particles well separated in rapidity. Evidence indicating long-range angular correlations include the appearance of the so-called near-side "ridge" in correlation functions and  $p_T$ dependent flow components ($v_N$). These features have been well measured at RHIC for Au+Au collisions at $\sqrt{s_{NN}} $= 200 GeV and at the LHC for Pb+Pb collisions at $\sqrt{s_{NN}}$ = 2.76 TeV. However, evidence of collectivity has recently been observed at the LHC in p+p high multiplicity collisions at $\sqrt{s_{NN}}$ = 7.0 TeV  and p+Pb collisions at $\sqrt{s_{NN}}$ = 5.02 TeV and then at RHIC in d+Au and $3^He$+Au collisions at $\sqrt{s_{NN}}$ = 200 GeV. In this talk, we present the recent PHENIX results for d+Au and $^3He$+Au collisions at $\sqrt{s_{NN}}$ = 200 GeV. Precise measurements of anisotropy help distinguish between theoretical models based on relativistic hydrodynamics or Color Glass Condensate (CGC). Comparisons of these measurements to various theoretical models are shown.
\end{abstract}


\end{frontmatter}


\section{Introduction}\label{intro}
Small collision systems have been considered too small to create hot and dense matter and exhibit signs of collectivity. However, this assumption has been challenged with the recent measurements of what appears to be collectivity in d+Au and p+Pb collisions at RHIC~\cite{Yi2014326} and the LHC~\cite{JHEP.10.1007}. In order to understand whether hot and dense matter can be created in small collision systems or not, detailed measurement indications of collectivity, such as long-range angular correlations, have been made by the PHENIX collaboration.

The observation of collectivity in matter can be a powerful indicator of fundamental properties in that matter. Collectivity means many discrete structures are interacting together to form a whole. In high energy heavy ion physics, collectivity means a medium is formed that can be described as a locally equilibrated system evolving hydrodynamically instead of a group of individually interacting constituents. 
Collectivity is measured by looking for long-range angular correlations in the spray of final state particles that come out of the collision. A key property of high energy heavy ion collisions is that information on the initial conditions will be carried through the medium evolution. Thus, an asymmetry in the initial conditions of the heavy ion collision is measurable in the final products.

A measurement of the azimuthal anisotropy is a way to quantify the extent of long-range angular correlation present in the medium evolution. 
One way to study the azimuthal anisotropy is to create a correlation function. The 2-particle correlation function method uses pairs of particles from the event in order to create a correlation function. For each each pair in an event, a $\Delta\phi$ value is obtained which makes up the signal $S(\Delta\phi,p_T)$. In order to correct for artificial correlations which would distort the distribution from detector effects or other sources, a mixed event background distribution $M(\Delta\phi,pT)$ is created. The correlation function can be defined as follows:

\begin{equation}\label{eqn:corr_func}
  C(\Delta,p_T) = \frac{S(\Delta\phi,p_T)}{M(\Delta\phi,p_T)}\frac{\int_{0}^{2\pi}M(\Delta\phi,p_T)d\Delta\phi)}{\int_0^{2\pi}S(\Delta\phi,p_T)d\Delta\phi)}
\end{equation}

Substantial variations in this $C(\Delta\phi,p_T)$ are usually seen as long-range angular correlations which can be attributed to collectivity.

In order to quantify the azimuthal anisotropy, $C(\Delta\phi,p_T)$ is Fourier transformed:
\begin{equation}\label{eqn:dndphi}
  C(\Delta\phi,p_T) \propto 1 + \sum_{n=1}2 v_{n}\cos(n[\phi(p_T)-\Psi_n]) 
\end{equation}

where $\Psi_n$ is the generalized event plane angle, $\phi$ is the azimuth of tracks from the event, and $v_n$ are flow coefficients. The measured $v_n$ averaged over a single event is defined as:
\begin{equation}\label{eqn:vn}
  v_n = \frac{\langle\cos(n[\phi-\Psi_n])\rangle}{Res(\Psi_n)}
\end{equation}

where $Res(\Psi_n)$ is the event plane resolution for each event. The $v_N$ that are reported in these proceedings are further averaged over each event.

The PHENIX collaboration is able to measure $v_n$ for charged particles in small collision systems as a function of transverse momentum $p_T$ (from 0.5 GeV to 3.0 GeV).
This proceeding will focus on recent measurements of $C(\Delta\phi,p_T)$ and $v_N$ in small collision systems.

\section{Experimental Setup}\label{expt}

The PHENIX detector~\cite{NIMA.499.469} has a high rate capability utilizing a fast DAQ 
and specialized triggers, high granularity detectors, and good momentum resolution and particle ID. Precise measurements of the event plane were carried out by forward PHENIX detectors such as the muon piston calorimeter (MPC), the beam-beam counter (BBC) and the forward silicon vertex detector (FVTX) (installed in 2012). Charged particle tracking at mid rapidity was done by the drift chamber (DCH), the pad chamber (PC1), and the vertex detector (VTX) (installed in 2011).

In 2008, the Relativistic Heavy Ion Collider (RHIC) at Brookhaven National Laboratory 
(BNL) produced $d$+Au collisions at $\sqrt{s_{NN}}$ = 200 GeV and in 2014, RHIC produced $^3He+Au$ collisions at $\sqrt{s_{NN}}$ = 200 GeV. The presence of a PHENIX high multiplicity trigger in 2014 allowed the collection of 7 times more 0-5\% $^3He+Au$ data than would normally be taken, which is vital for obtaining enough relevant data.
Results from these data sets will be presented in this proceeding.
\section{Results}\label{result}
The PHENIX collaboration measured the azimuthal correlation functions $C(\Delta\phi,p_T)$  and flow coefficients $v_n$ in d+Au ~\cite{PhysRevLett.114.192301} and $^3He$+Au ~\cite{arxiv.1507.06273} collisions at $\sqrt{s_{NN}}$ = 200 GeV . 

\begin{figure}
\includegraphics[width=1.0\linewidth]{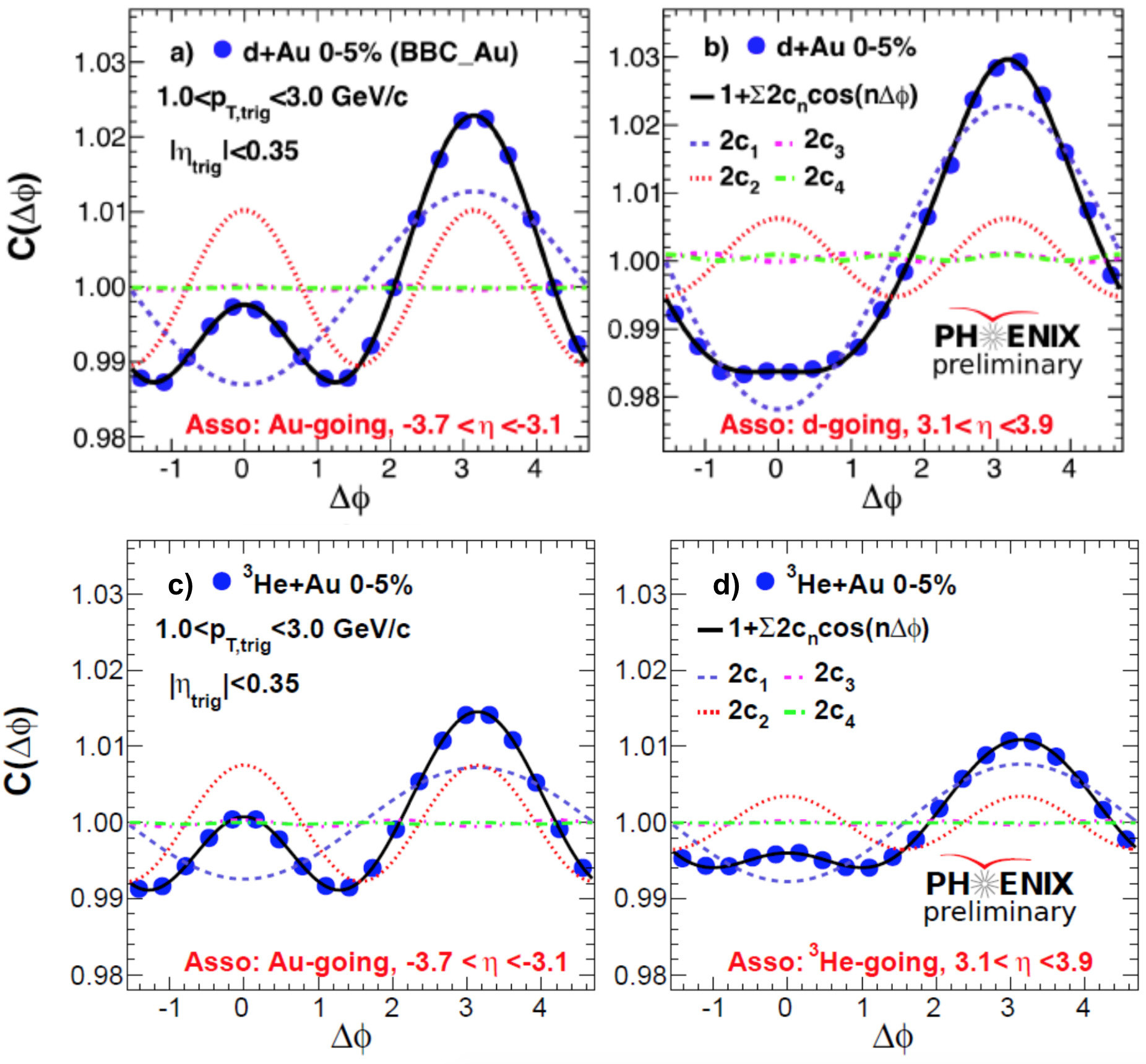}
\caption{\label{fig:ridge_heau_dau}  Azimuthal correlation functions $C(\Delta\phi,p_T)$ for track-BBC PMT pairs in different collisions systems and with different selections of associated BBC hits. The four correlation functions (blue points) are from a) d+Au 0-5$\%$ centrality collisions with associated tracks in the Au-going direction, b) d+Au 0-5$\%$ centrality collisions with associated tracks in the d-going direction, c) $^3He$+Au 0-5$\%$ centrality collisions with associated tracks in the Au-going direction, and d) $^3He$+Au 0-5$\%$ centrality collisions with associated tracks in the $^3He$-going direction.}
\end{figure}
Figure~\ref{fig:ridge_heau_dau} show the correlation functions. The observed near-side ridge (the bump in $C(\Delta\phi,p_T)$ at around $\Delta\phi$ = 0) is much larger for both the d+Au and $^3He$+Au collision systems in the Au-going direction than the other direction. This represents an enhancement in the long-range correlation observed at backward rapidity as opposed to forward rapidity. Another significant feature is the difference in the near-side ridge in the Figure~\ref{fig:ridge_heau_dau} b) and d). The almost total lack of a ridge in d-going direction is highlighted when compared to the ridge seen in the $^3He$-going direction.
 
 \begin{figure}
\includegraphics[width=1.0\linewidth]{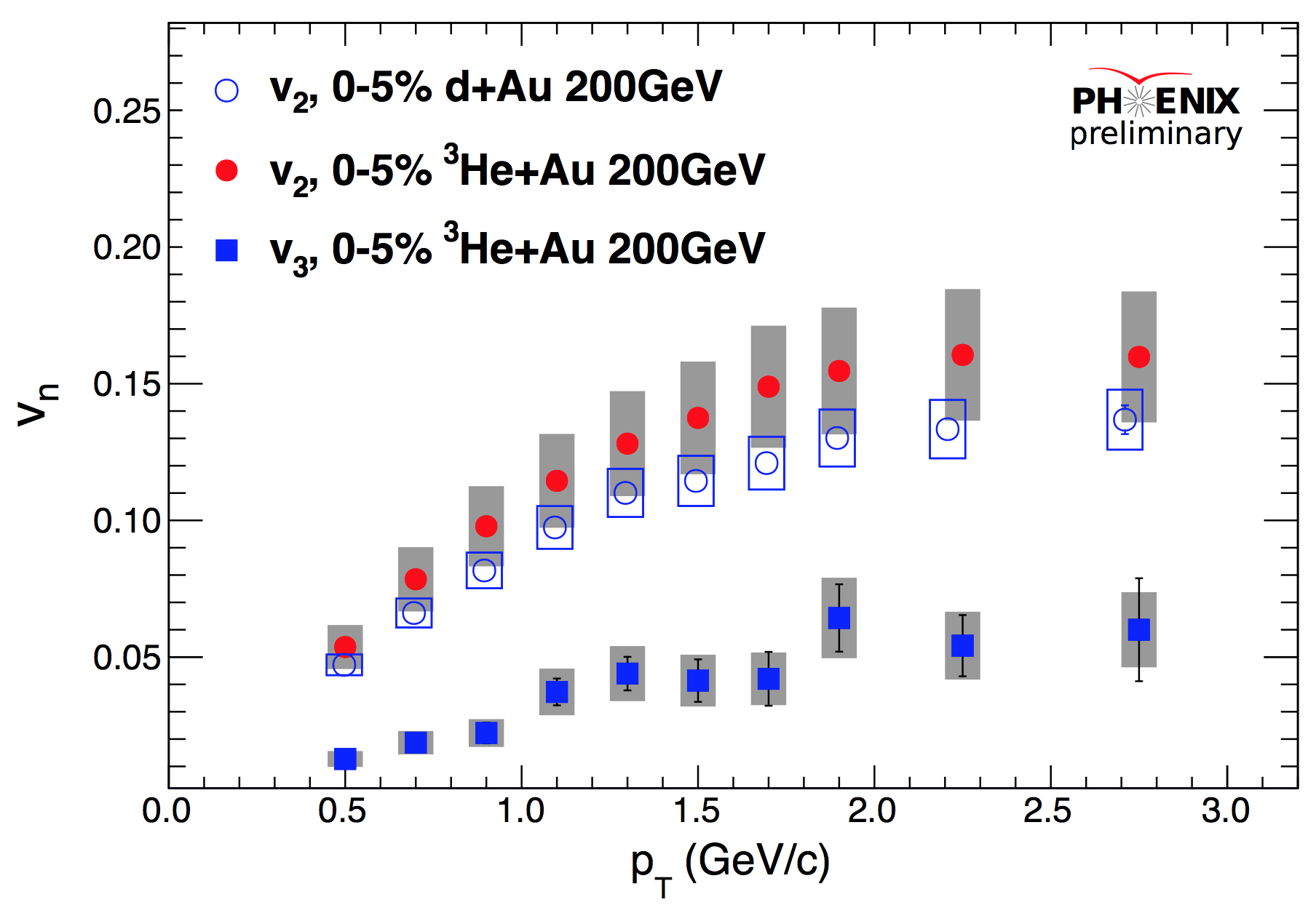}
\caption{\label{fig:vn_dau_heau}  Second and third order flow coefficients $v_2 $and $v_3$ $p_T$ dependence as measured by PHENIX in 0-5\% centrality $^3He$+Au collisions and d+Au 0-5\% collisions at $\sqrt{s_{NN}}$ = 200 GeV.}
\end{figure}

Figure~\ref{fig:vn_dau_heau} shows the $v_2$ and $v_3$ as a function of $p_T$ for $^3He$+Au and the $v_2$ as a function of $p_T$ for d+Au events. A clear $p_T$ dependence is seen for $v_2$ in both the collision systems. The d+Au $v_2$ is comparable to the $^3He$+Au $v_2$ in shape and in overall scale. Another thing to note is the substantial $v_3$ signal that is measured. This measurement is consistent with the interpretation that the intrinsic initial geometry of  $^3He$+Au 0-5\% centrality collisions is being propagated through the medium evolution to the final state particles. Detailed studies of Glauber + hydrodynamic simulations of $^3He$+Au 0-5\% events at $\sqrt{s_{NN}}$ = 200 GeV to understand how the initial geometry is propagated to the final state particles reveal that the initial triangular orientation becomes inverted as the medium evolves~\cite{PhysRevLett.113.112301}.

 \begin{figure}
\includegraphics[width=1.0\linewidth]{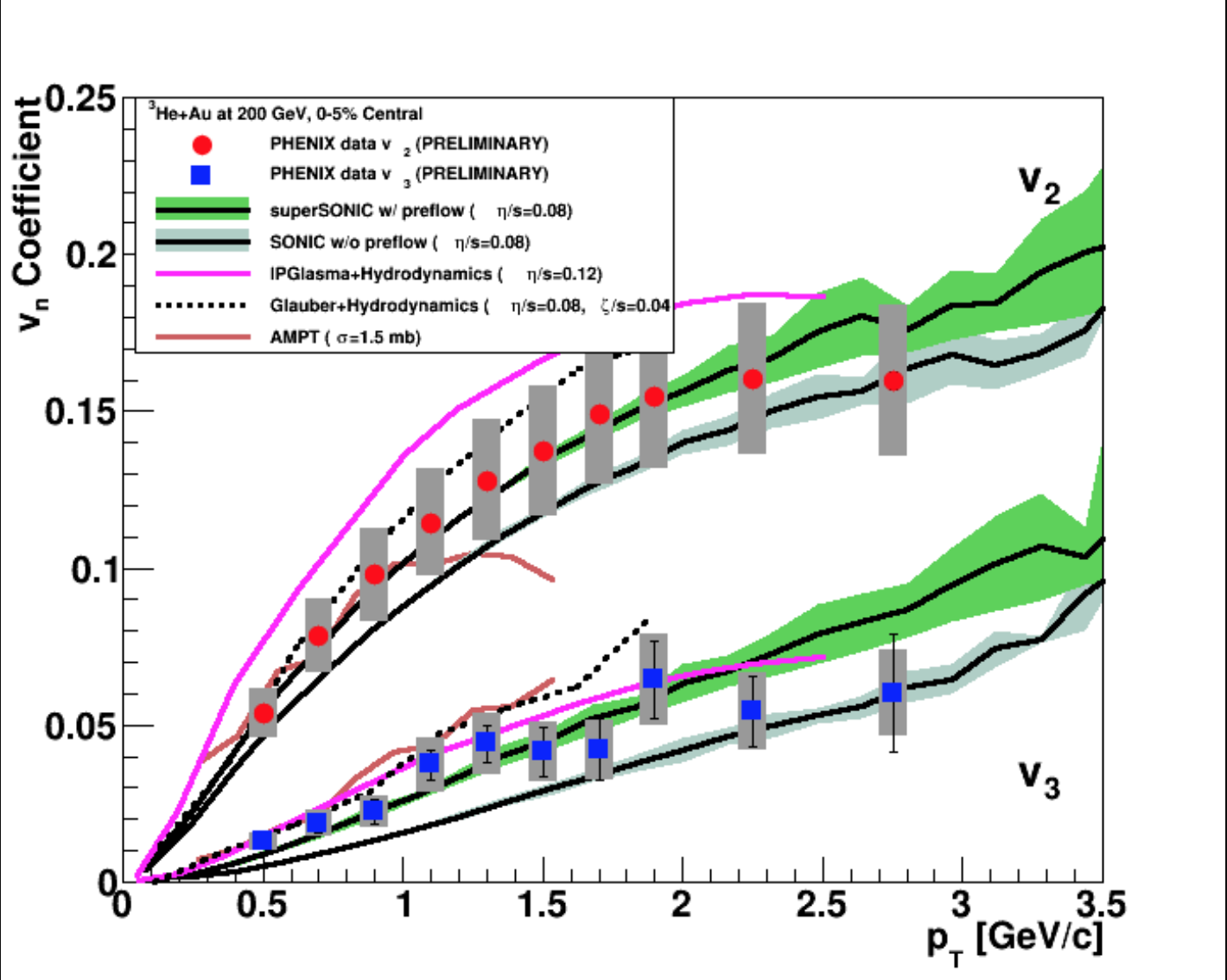}
\caption{\label{fig:vn_heau_w_theo}  Second and third flow coefficients $v_2$ and $v_3$ $p_T$ dependence as measured by PHENIX  in 0-5\% centrality $^3He$+Au collisions $\sqrt{s_{NN}}$ = 200 GeV with several theory curves plotted. Details of the SONIC models can be seen here~\cite{arXiv:1502.04745} and the AMPT can be seen here~\cite{arXiv:1501.06880}.}
\end{figure}

Figure~\ref{fig:vn_heau_w_theo} shows same $v_2$ and $v_3$ measurements as a function of $p_T$ for 0-5\% $^3He$+Au collisions at 200 GeV as seen in figure ~\ref{fig:vn_dau_heau}, except with a multitude of theoretical model curves plotted too. The best agreement with the points comes from the superSONIC with preflow model which uses Glauber, hydrodynamics, as well as pre-equilibrium flow and hadronic cascade afterburner ~\cite{arXiv:1502.04745}. One of the more common theoretical models, Glauber initial conditions + hydrodynamics evolution, shows some agreement with the measured points up to 2 GeV. The IPGlasma initial conditions + hydrodynamic evolution model tends to over-predict the $v_2$ points in contrast to its agreement with the $v_3$ points. The AMPT (a multi-phase transport) model curve also shows agreement with the data points ~\cite{arXiv:1501.06880}. The fact that these models have the assumption of collectivity and flow, coupled with the good agreement of the models' theory curves to the data implies that collectivity and flow may explain what is being observed in small collision systems.

\section{Summary and Outlook}\label{summ}

The PHENIX collaboration has measured correlation functions and flow coefficients in small collision systems: d+Au and $^3He$+Au 0-5\% centrality collisions at $\sqrt{s_{NN}}$ = 200 GeV. 

Observed in the correlation functions was a clear near-side ridge in both the Au-going direction and the $^3He$-going direction in $^3He$+Au 0-5\% collisions at $\sqrt{s_{NN}}$ = 200 GeV and a clear near-side ridge in the Au-going direction in d+Au 0-5\% collisions at $\sqrt{s_{NN}}$ = 200 GeV. The presence of this near-side ridge is significant because it may indicate collectivity.

Observed in $v_2$ in $^3He$+Au and d+Au 0-5\% collisions at $\sqrt{s_{NN}}$ = 200 GeV is a clear $p_T$ dependence. Observed in the $^3He$+Au collisions is a substantial $v_3$. Comparison to various models which use medium evolution reveal good agreement and imply that something that resembles collectivity is being observed in these small collision systems.

The PHENIX collaboration is working to continue the search for collectivity in small QCD systems by looking for a near-side ridge in high multiplicity p+p collisions, measuring $v_N$ in p+Au collisions, and measuring the pseudorapidity dependence of $v_2$ in $^3He$+Au and p+Au collisions. 




\section{References}
\bibliographystyle{elsarticle-num}
\bibliography{koblesky_t}

\end{document}